\documentclass[%
 reprint,
superscriptaddress,
 amsmath,amssymb,
 aps,
prc,
floatfix,
]{revtex4-1}
\usepackage{multirow}
\usepackage{graphicx}
\usepackage{dcolumn}
\usepackage{bm}
\usepackage{array}
\usepackage{hyperref}
\DeclareUnicodeCharacter{200E}{ }
\usepackage[mathlines]{lineno}

\begin{document}

\preprint{APS/123-QED}

\title{Potential absence of observed $\pi^2$ linear-chain structures in $^{14}$O via $^{10}$C($\alpha,\alpha$) resonant scattering}
		\author{J.~Bishop}
  \email{j.bishop.2@bham.ac.uk}
     \affiliation{School of Physics and Astronomy, University of Birmingham, Edgbaston, Birmingham, B15 2TT, United Kingdom}
     
    \author{A.~Hollands}
     \affiliation{School of Physics and Astronomy, University of Birmingham, Edgbaston, Birmingham, B15 2TT, United Kingdom}
	    \author{Tz.~Kokolova}
     \affiliation{School of Physics and Astronomy, University of Birmingham, Edgbaston, Birmingham, B15 2TT, United Kingdom}
         	\author{G.V.~Rogachev}
		\affiliation{Cyclotron Institute, Texas A\&M University, College Station, TX 77843, USA}
		\affiliation{Department of Physics \& Astronomy, Texas A\&M University, College Station, TX 77843, USA}
		\affiliation{Nuclear Solutions Institute, Texas A\&M University, College Station, TX 77843, USA}
         \author{C.~Wheldon}
     \affiliation{School of Physics and Astronomy, University of Birmingham, Edgbaston, Birmingham, B15 2TT, United Kingdom}
\author{E.~Aboud}
          \altaffiliation{Current Address: Lawrence Livermore National Laboratory, Livermore, California 94550, USA}  
		\affiliation{Cyclotron Institute, Texas A\&M University, College Station, TX 77843, USA}
		\affiliation{Department of Physics \& Astronomy, Texas A\&M University, College Station, TX 77843, USA}
		\author{S.~Ahn}
          \altaffiliation{Current Address: Center for Exotic Nuclear Studies, Institute for Basic Science, 34126 Daejeon, South Korea}
  		\affiliation{Cyclotron Institute, Texas A\&M University, College Station, TX 77843, USA}

		\author{M.~Barbui}
		\affiliation{Cyclotron Institute, Texas A\&M University, College Station, TX 77843, USA}
         \author{N.~Curtis}
     \affiliation{School of Physics and Astronomy, University of Birmingham, Edgbaston, Birmingham, B15 2TT, United Kingdom}
		\author{J.~Hooker}
          \altaffiliation{Current Address: Los Alamos National Laboratory, Los Alamos, New Mexico 87545, USA}
		\affiliation{Cyclotron Institute, Texas A\&M University, College Station, TX 77843, USA}
		\affiliation{Department of Physics \& Astronomy, Texas A\&M University, College Station, TX 77843, USA}
		\author{C.~Hunt}
\altaffiliation{Current Address: Facility for Rare Isotope Beams, Michigan State University, East Lansing, Michigan 48824, USA}

		\affiliation{Cyclotron Institute, Texas A\&M University, College Station, TX 77843, USA}
		\affiliation{Department of Physics \& Astronomy, Texas A\&M University, College Station, TX 77843, USA}
		\author{H.~Jayatissa}
          \altaffiliation{Current Address: Los Alamos National Laboratory, Los Alamos, New Mexico 87545, USA}
		\affiliation{Cyclotron Institute, Texas A\&M University, College Station, TX 77843, USA}
		\affiliation{Department of Physics \& Astronomy, Texas A\&M University, College Station, TX 77843, USA}
		\author{E.~Koshchiy}
        \altaffiliation{Deceased}
		\affiliation{Cyclotron Institute, Texas A\&M University, College Station, TX 77843, USA}	

         \author{S.~Pirrie}
     \affiliation{School of Physics and Astronomy, University of Birmingham, Edgbaston, Birmingham, B15 2TT, United Kingdom}
		\author{B.T.~Roeder}
		\affiliation{Cyclotron Institute, Texas A\&M University, College Station, TX 77843, USA}
		\author{A. Saastamoinen}
		\affiliation{Cyclotron Institute, Texas A\&M University, College Station, TX 77843, USA}
		\author{S.~Upadhyayula}
          \altaffiliation{Current Address: Lawrence Livermore National Laboratory, Livermore, California 94550, USA}  
		\affiliation{Cyclotron Institute, Texas A\&M University, College Station, TX 77843, USA}
		\affiliation{Department of Physics \& Astronomy, Texas A\&M University, College Station, TX 77843, USA}

\email{j.bishop.2@bham.ac.uk}

\date{\today}

\begin{abstract}
\begin{description}
\item[Background]
The preference for light nuclear systems to coagulate into $\alpha$-particle clusters has been well-studied. The possibility of a linear chain configuration of $\alpha$-particles would allow for a new way to study this phenomenon.
\item[Purpose]
A rotational band of states in $^{14}$C has been claimed showing a $\pi^2$ linear chain structure. The mirror system, $^{14}$O, has been studied here to examine how this linear chain structure is affected by replacing the valence neutrons with protons.
\item[Method]
A beam of $^{10}$C was incident into a chamber filled with He:CO$_2$ gas with the tracks recorded inside the TexAT Time Projection Chamber and the recoil $\alpha$-particles detected by a silicon detector array to measure the $^{10}\mathrm{C}(\alpha,\alpha)$ cross section. 
\item[Results]
The experimental cross section was compared with previous studies and fit using R-Matrix theory with the previously-observed $^{14}$O states being transformed to the $^{14}$C using mirror symmetry. The measured cross section does not replicate the claimed states, with the predicted cross section exceeding that observed at several energies and angles.
\item[Conclusion]
A series of possibilities are highlighted with the most likely being that the originally-seen $^{14}$C states did not constitute a $\pi^2$ rotational band with a potentially incorrect spin assignment due to the limitations of the angular correlation method with non-zero spin particles. The work highlights the difficulties in measuring broad resonances corresponding to a linear chain state in a high level density.
\end{description}
\end{abstract}

\pacs{Valid PACS appear here}
\maketitle

\section{\label{sec:Introduction}Introduction}

The importance of $\alpha$-cluster phenomena, which are prevalent across light nuclei \cite{Ikeda}, is becoming increasingly apparent in the context of near-threshold states in open quantum systems \cite{OQS,Oertzen,B11}. Traditionally, the strongest evidence of $\alpha$-clusters exists in $\alpha$-conjugate nuclei with the $^{8}$Be ground state and the Hoyle state ($J^{\pi}=0_{2}^{+}$ at 7.65~MeV) in $^{12}$C \cite{Otsuka, Hoyle, HoyleFreer} being two textbook examples. These $\alpha$-conjugate nuclei have had success in being described in terms of rigid geometric cluster configurations such as the $D_{3h}$ model which well reproduces a variety of states in $^{12}$C \cite{D3h_t, D3h_e}. Extending such geometric configurations to non $\alpha$-conjugate nuclei has also been an area of interest for the past few decades. In particular, the question of how the addition of valence neutrons or protons affects the underlying $\alpha$-cluster structure is of significant interest and offers a great test for \emph{ab initio} models. There has been a great amount of interest in studying the possibility of a 3 $\alpha$-particle linear chain state in A$>$12 systems. For $^{12}$C, it has been long understood that the Hoyle state, by virtue of its moment of inertia, cannot be a linear chain configuration \cite{LC1,LC2,LC3} and this linear configuration is unstable. The addition of valence protons or neutrons modifies this for heavier systems, and the additional nucleons can make the system stiffer and therefore able to support such a state \cite{Suhara}. Recently, evidence of the 3 $\alpha$-particle linear chain state was purported in $^{14}$C via the $^{14}$C($p,p'$) inelastic scattering reaction \cite{Han}. Observation of the breakup of states into $^{10}$Be + $\alpha$ allowed for a series of states to be reconstructed using invariant mass. A selection of these states were suggested as corresponding to a rotational band of $J^{\pi}$ = 0$^{+}$, 2$^{+}$, and $4^{+}$ using spin-parity analysis with the angular correlation method \cite{AngCorr}. Given the large moment of inertia of these states, which agree reasonably well with AMD (Antisymmetrized Molecular Dynamics) calculations for the linear chain configuration \cite{AMD3} by Baba and Kimura, these states were given as evidence of the linear chain structure in $^{14}$C with two p-wave neutrons around a 3$\alpha$ linear chain. \par
Examination of states in the mirror system where the valence neutrons are instead valence protons allows for a great check of the mirror symmetries of such a system in terms of the Thomas-Ehrmann shift (which can provide information about the spatial extent of the valence neutrons) and also the robustness of mirror symmetry where the reduced width for mirror states should be similar. Observation of the same set of linear chain states in the mirror system will also provide proof or possible refutation of the observations in $^{14}$C. \par
Therefore, an experiment was performed to search for $\alpha$-cluster structures in $^{14}$O via the $^{10}$C($\alpha,\alpha)$ elastic scattering reaction.

\section{\label{Sec:exp}Experimental Setup}

\begin{figure}
\centerline{\includegraphics[width=0.5\textwidth]{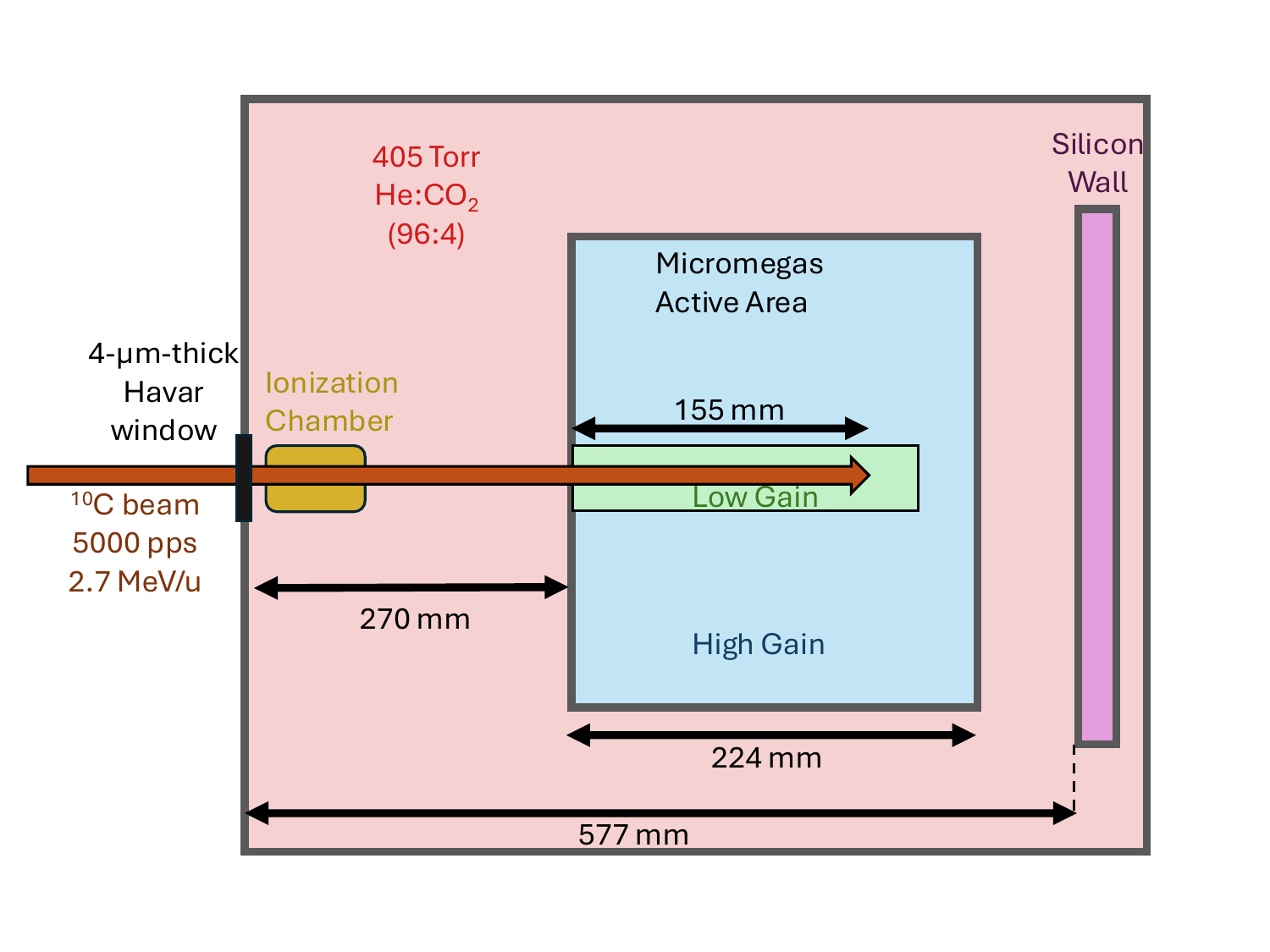}}
\centerline{\includegraphics[width=0.5\textwidth]{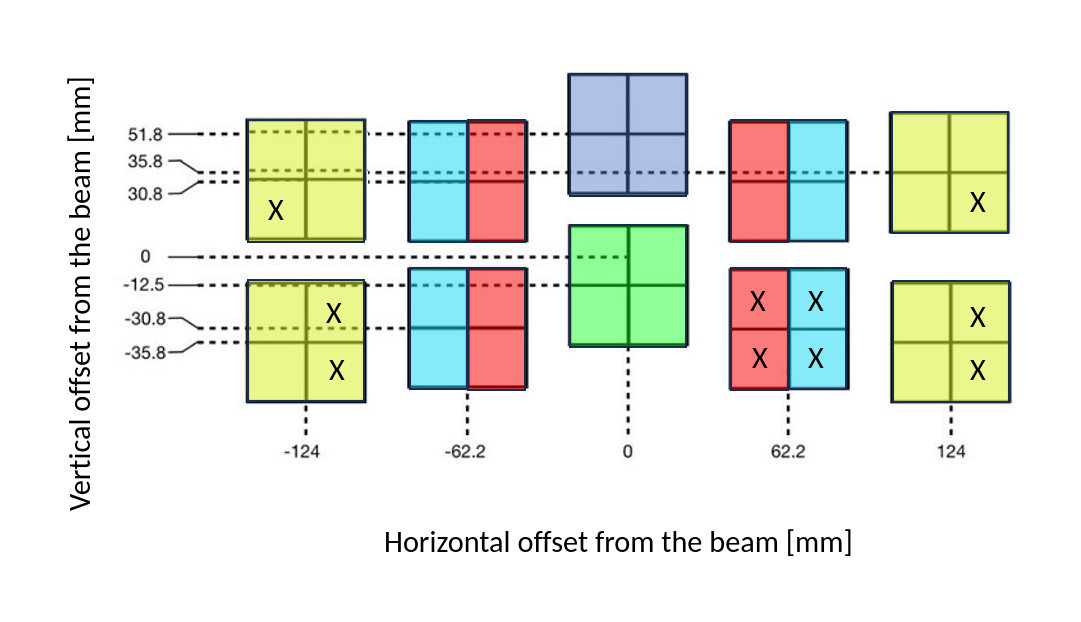}}
\caption{\label{fig:detectors} (Top) Schematic of the detector setup with distances between the window, active region of the TPC and the silicon forward wall denoted. (Bottom) Layout of the forward wall of detectors showing their horizontal and vertical offset with respect to the beam. Quadrants which were not used for data analysis (primarily due to detector issues) are indicated with an ``X". The five colored regions (color online) show the separation of different detector quadrants to form five different angular ranges.}
\end{figure}

In order to measure the $^{10}$C($\alpha,\alpha)$ reaction, a radioactive beam of $^{10}$C (half life 19.3 s) was produced at the Cyclotron Institute, Texas A\&M University using the K500 Cyclotron to provide a 7 MeV/nucleon $^{10}$B beam which was incident on a LN$_2$-cooled H$_2$ target at 870 Torr. Via the $^{10}$B($p,n)^{10}$C reaction, a secondary radioactive beam of $^{10}$C was separated from contaminants using the Momentum Achromatic Recoil Separator (known as MARS). The beam then entered the chamber at an energy of around 2.7~MeV/nucleon with a typical intensity of 5000 pps. In order to measure the reaction, the Texas Active Target Time Projection Chamber (TexAT TPC) \cite{TexAT} was used and was filled with 405 Torr of He:CO$_2$ (96:4). Using the Thick Target in Inverse Kinematics (TTIK) approach \cite{TTIK}, the elastic scattering cross section was measured from an energy of around 6.5 MeV in the center of mass down to 2 MeV with the beam stopping inside the chamber (155 mm into the active area of the TPC). The tracks of the charged particles were measured using a segmented Micromegas detector which had a higher gas gain away from the central beam region in order to identify better the tracks from the scattered $\alpha$ particles. The active region of the TPC starts 270 mm from the 4-$\mu$m-thick Havar entrance window. To measure the total energy of the $\alpha$ particles with a good energy resolution, an array of MSQ-25 silicon detectors (5 x 5 cm$^2$ quad detectors) were placed downstream to pickup the recoil particles. The detectors, with thicknesses ranging from 625-700 $\mu$m, were sufficiently thick to fully stop the $^{4}$He ions of interest. The acquisition was then triggered by a signal in the forward silicon wall. This setup is identical to that covered in previous work to study $^{14}$O$(\alpha,\alpha)$ \cite{Marina} having run sequentially with the current work. The silicon wall is shown in Fig.~\ref{fig:detectors} and is 307 mm from the start of the active region of the TPC. Given here in the lab frame: the central `zero-degree' detector (green online, 0-6$^{\circ}$) comprises all four quadrants, the central `upper' detector (dark blue online, 1-16$^{\circ}$) also comprises all four quadrants. The inner half of the next detectors from the center form the `inner' angular region (red online, 2-17$^{\circ}$) with the outer half of those same detectors forming the `middle' angular region (cyan online, 6-20$^{\circ}$). Finally, all of the quadrants of the outermost detectors form the `outer' angular region (gold online, 11-29$^{\circ}$). 

\section{Data Analysis}

\begin{figure}
\centerline{\includegraphics[width=0.5\textwidth]{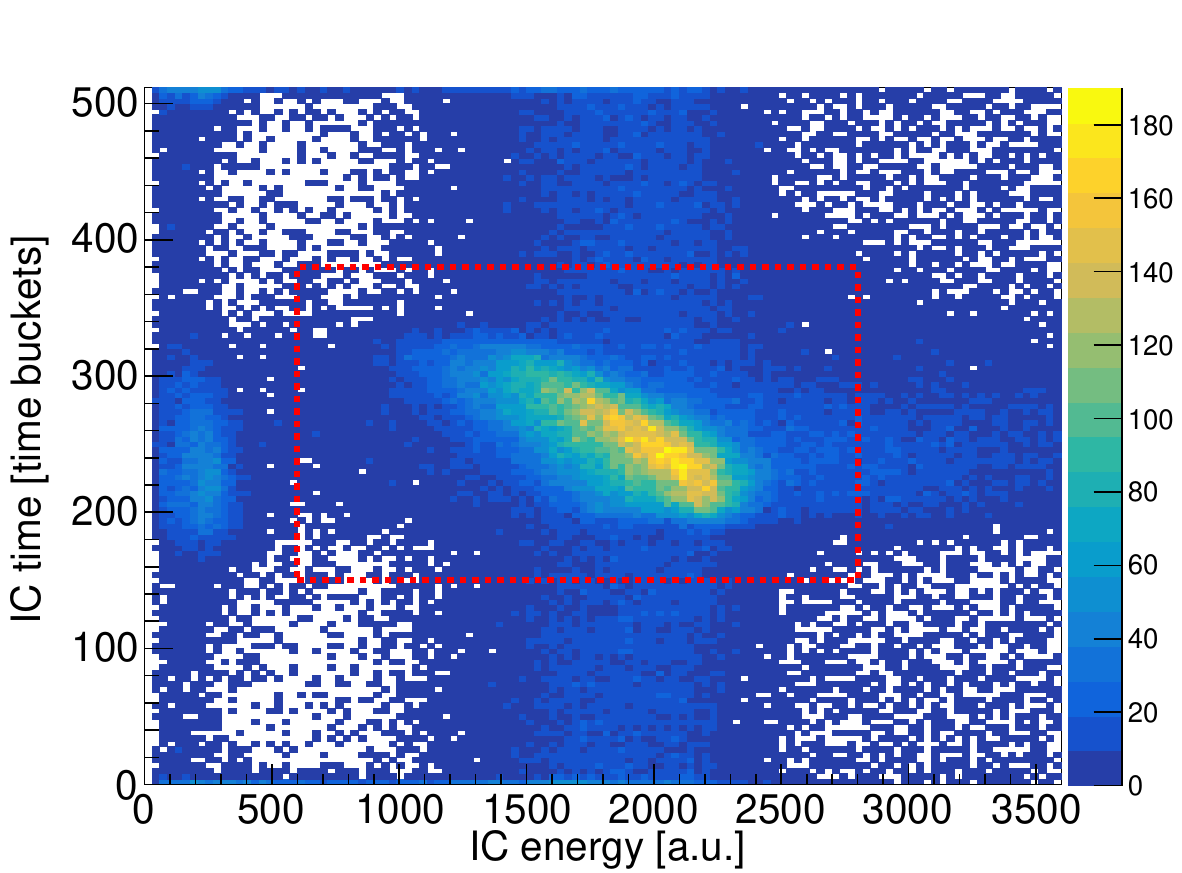}}
\caption{\label{fig:IC}Ionization chamber (IC) energy versus time plot showing the $^{10}$C coincident peak within the 2D gate (red dotted line). A small contaminant ($^{7}$Be) can be seen around an energy of 200 (arbitrary units) that is hence excluded.}
\end{figure}

While the radioactive beam was of very high purity ($>$99\%), the signal from an ionization chamber (IC) at the entrance to the chamber was used to verify the incoming beam was $^{10}$C. The particle ID plot for this is shown in Fig.~\ref{fig:IC} where the time of the IC signal against the silicon trigger is plotted against the amplitude. In this way, random coincidences originating from $\beta^+$ firing the silicon detectors from the decay of the $^{10}$C stopped inside the chamber could be eliminated, as well as lighter contaminants ($^{7}$Be) that did not stop in the gas and struck the zero degree detectors.\par

\begin{figure}
\centerline{\includegraphics[width=0.5\textwidth]{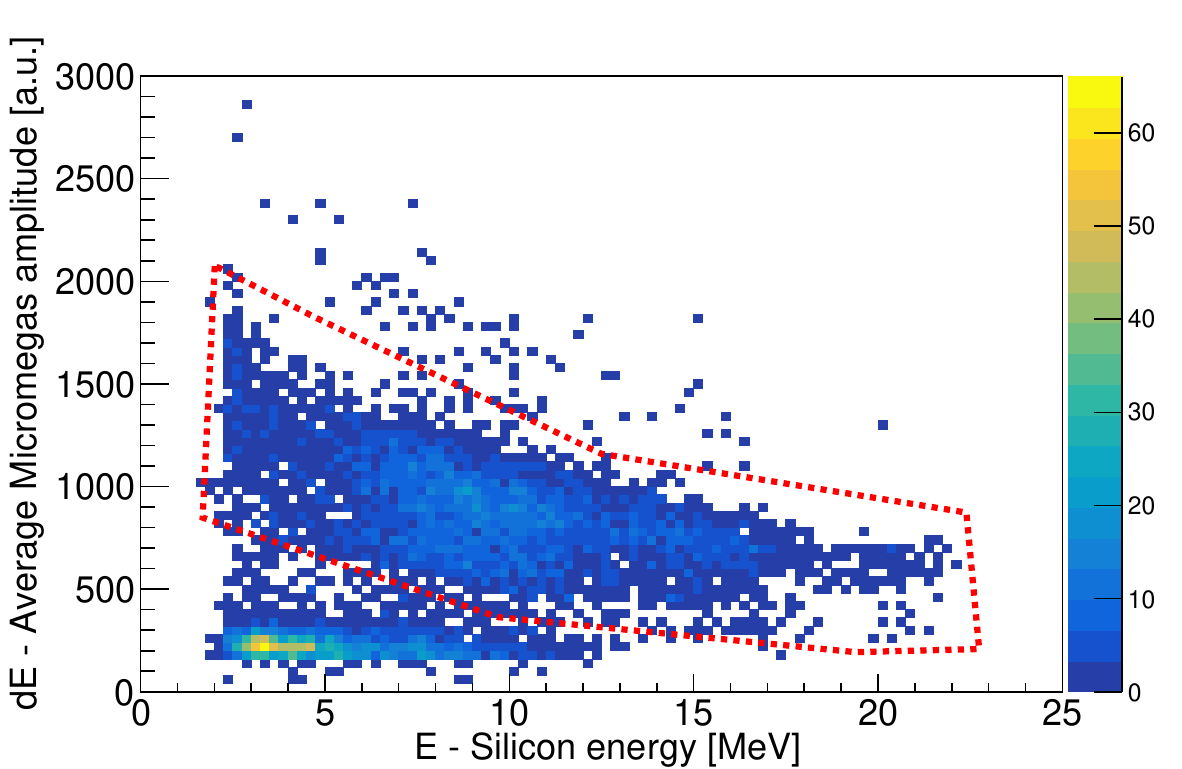}}
\caption{\label{fig:hdEE}Identification of the charge of the recoil particle detected in the silicon detectors. The average energy loss in the Micromegas is compared to the energy deposited in the silicon detector. The band for Z=2 can clearly be seen against the Z=1 band as well as a weaker Z=3 band. The gate to select Z=2 is overlaid by a red dotted line.}
\end{figure}

Given the incoming ion was now identified as $^{10}$C, the recoil particle hitting the silicon detector was then identified using the energy loss in the Micromegas detector (dE) vs the energy in the silicon (E). The band corresponding to Z=2 (where the resolution is insufficient to properly exclude $^{3}$He) is shown in Fig.~\ref{fig:hdEE} for a subset of runs. There were small gas gain changes during the experiment which were accounted for by modifications to the gate for Z=2. \par

\begin{figure}
\centerline{\includegraphics[width=0.5\textwidth]{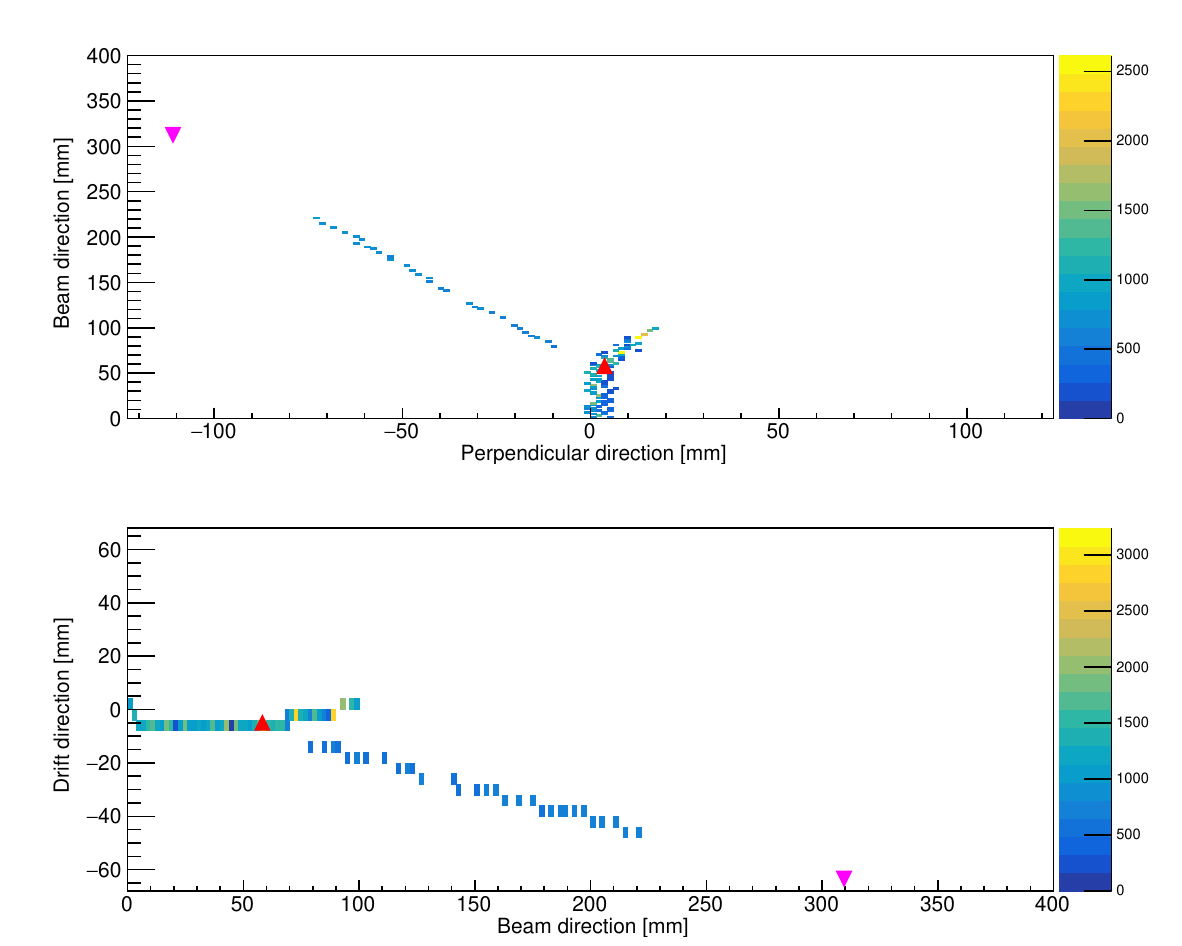}}
\caption{\label{fig:insideMMtrack}Top down (top panel) and side view (bottom panel) of a single $^{10}\mathrm{C}(\alpha,\alpha)$ scattering event with the interaction vertex inside of the Micromegas active region. The upright red triangle points (color online) show the fitted interaction vertex from the RANSChiSM algorithm and the silicon hit location is shown as an upside-down magenta triangle. The color scale denotes the energy deposition in the Micromegas with the lower dE/dx of the $\alpha$ (travelling left and down relative to the beam) being apparent.}
\end{figure}

In order to separate the elastic and inelastic channels, the tracks from the TPC were used to identify the interaction vertex which fell into two categories. Firstly, for events which happened inside the active area of the Micromegas, the tracks were fitted using a line for the incoming beam which extended from the entrance window to the interaction vertex, and two lines originating from the vertex for the light and heavy recoil. An example of such an event is shown in Fig.~\ref{fig:insideMMtrack}. The second category of event is those where the interaction vertex is not inside the active region of the Micromegas. In this case, the heavy recoil fragment may not enter the active region and the only track corresponds to that of the light particle. Therefore, the track for the light particle is fitted and the vertex location is chosen by a cubic region corresponding to the approximate extent of the beam profile. This essentially involves looking at the overlap between the light particle track and the z beam-axis but also allows for a small perpendicular offset due to the finite size of the beam. The fitting for the track uses the RanSChiSM technique used previously in other experiments using TexAT \cite{bishop2024cluster} and has been shown to work remarkably well, even with up to four tracks. \par

\begin{figure}
\centerline{\includegraphics[width=0.5\textwidth]{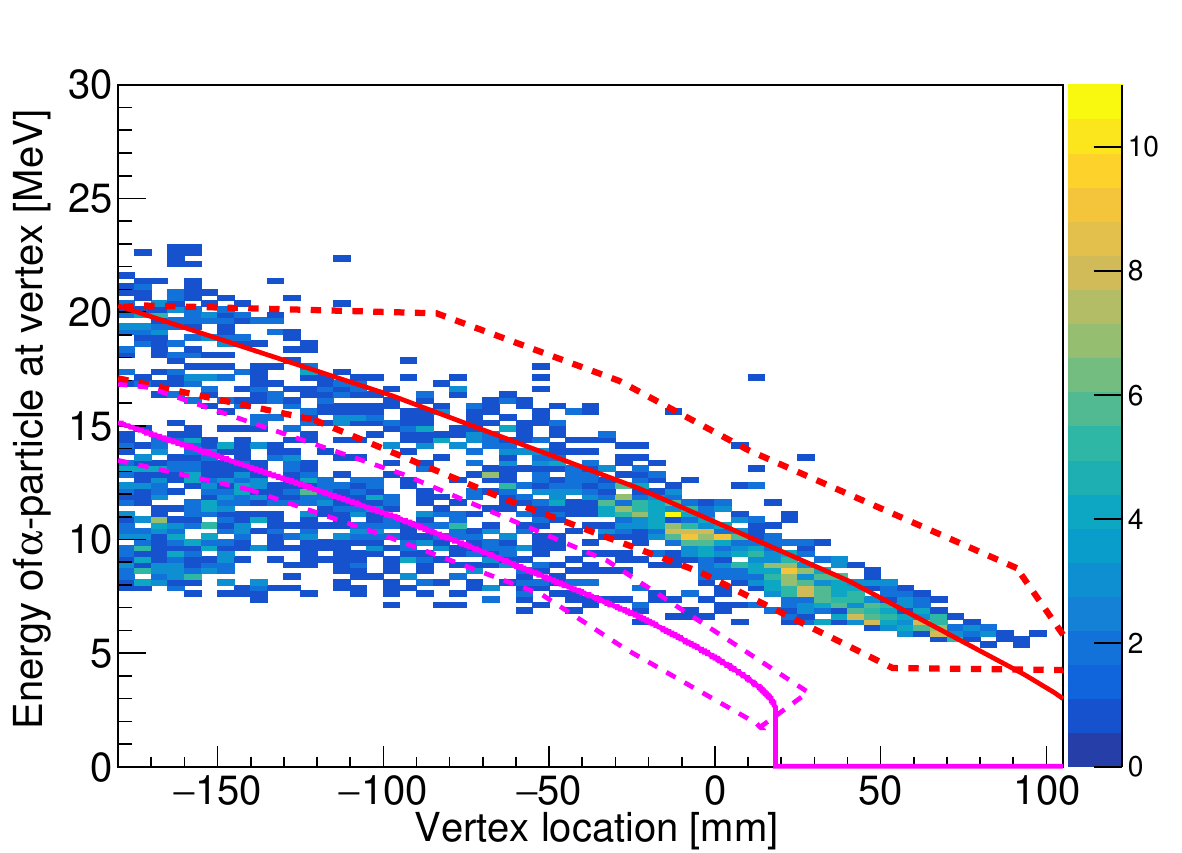}}
\caption{\label{fig:hEvertex}Channel selection plot for the elastic channel (red) versus the first-excited state (magenta) identified by analysing the interaction vertex location inside TexAT (relative to the start of the Micromegas) versus the energy of the $\alpha$-particle at the vertex for the outer detectors. An energy loss correction through the gas is applied to reconstruct the energy at the vertex. The solid lines show the expected behavior from kinematics for an $\alpha$-particle scattering at zero degrees with the dashed lines showing the 2D cuts that were applied to the data.}
\end{figure}

\begin{figure}
\centerline{\includegraphics[width=0.5\textwidth]{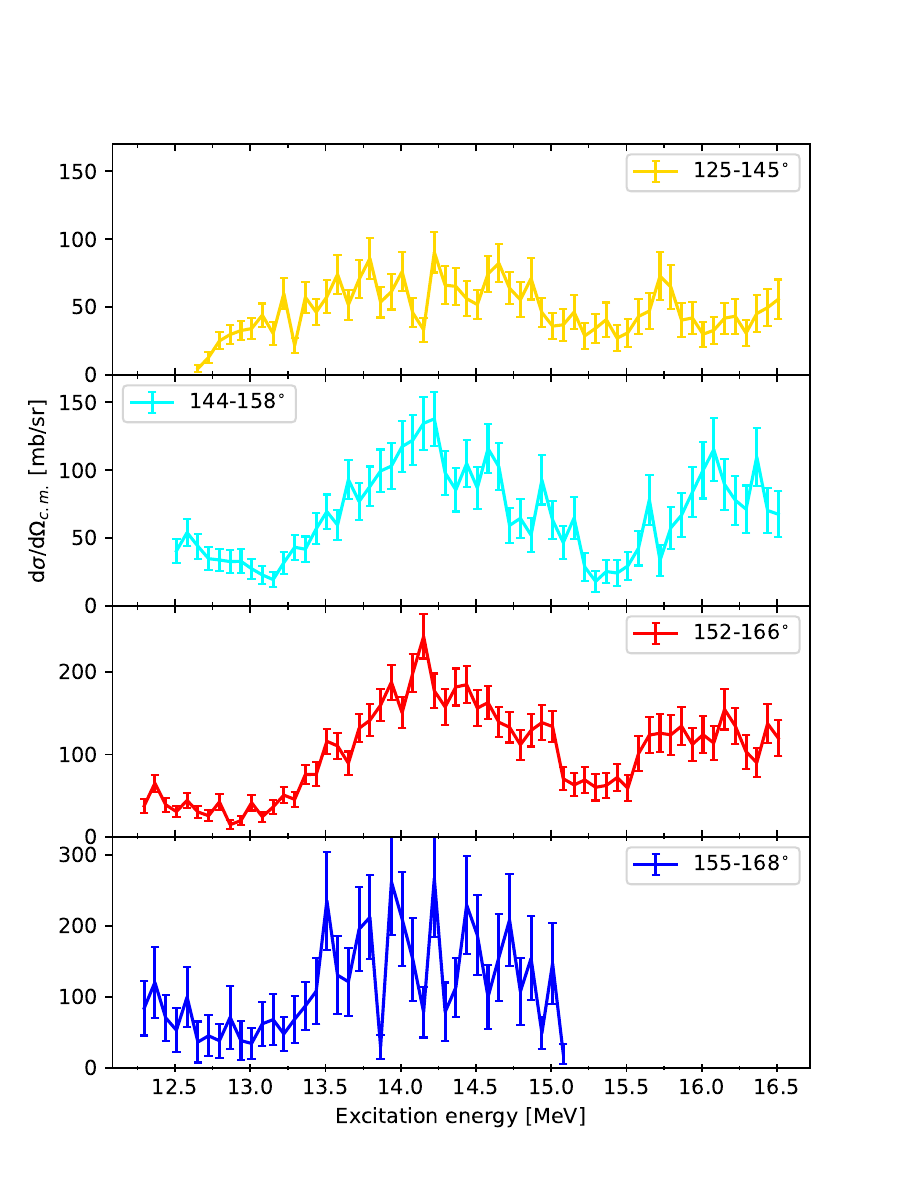}}
\caption{\label{fig:data}Measured center of mass differential cross section for $^{10}$C$(\alpha,\alpha)$ for the four different detector angles for the current work as shown in Fig.~\ref{fig:detectors}.}
\end{figure}

Once the vertex location is identified, a selection between the elastic and inelastic channels is made by examining the vertex location against the energy of the $^{4}$He recoil: $\alpha$-particles originating from the same vertex location (and the same excitation energy in $^{14}$O) from the inelastic channel will have a smaller $\alpha$-particle energy. The selection plot is shown in Fig.~\ref{fig:hEvertex} with the cuts for the elastic and first excited state inelastic channel cuts. The center of mass energy of the interaction is then calculated by correcting for the energy loss of the $^{4}$He through the gas (which has a well-known distance due to the track) and the scattering angle for the two channels. By separating the silicon quadrants into different regions, five data sets at different angles were generated. The data covering the zero-degree detector were omitted at this stage from the analysis due to the similarity in angle of a forward-scattered $\alpha$-particle with the beam. This meant that the vertex was very difficult to reliably reconstruct and these data were seen to be unreliable. Similarly for the `upper' detector, the data were only used in the region where the vertex was located inside the active region of the TPC. For this `upper' detector, the heavy recoil could be sufficiently well separated from the light recoil in the drift direction (vertically) --- this corresponds to a cut of $E_{c.m}$ $<$ 5 MeV. As the vertex location is not localized due to an extended gas target, the angular range covered by each of the quadrants is a function of the center of mass energy. Using the total beam fluence of 5.6 $\times$ 10$^{9}$ $^{10}$C ions, the differential cross sections were calculated and input into AZURE2 \cite{AZURE} to perform an R-Matrix fit including a Gaussian smearing effect to account for the experimental resolution ($\sigma E_{c.m.} = 80$ keV). Given the energy of the inelastic channel threshold, the statistics for this channel were extremely low and were not included in the final R-Matrix fit, but the channel selection was very important in order to remove the inelastic contribution. The calculated c.m.  differential cross section for the four angles included in the final R-Matrix analysis is shown in Fig.~\ref{fig:data}. \par

It is immediately obvious that there are no strong narrow resonances apparent and the cross section shows two very broad features with a width of $>$1 MeV. For $^{10}$C$\bigotimes \alpha$, one may calculate the Wigner limit which gives a maximum expected reduced width for a fully $\alpha$-clustered state. For this system, the Wigner limit is $\gamma^2_W$ = 800 keV (with $R_0 = 1.4$ fm). To get a maximum width, one can convert using $\Gamma = 2 P_{l}(E) \gamma^2$ with $P_l(E)$ being the penetrability factor for angular momentum $l$. At an excitation energy of 14.5 MeV, the largest penetrability factor ($l=0$) is 2.29. Therefore, the largest $\Gamma_{\alpha}$ one may achieve is 3.66 MeV. Realistically, for these light systems, a state is well-clustered with around 10$\%$ or more of the Wigner limit and does not often significantly exceed this value \cite{Wignerclustered}. Therefore these broad shapes are interference between the hard sphere scattering amplitude with multiple resonances and a small number of broad resonances that cannot be disentangled. The latter makes a complete and unambiguous R-Matrix fit extremely challenging but is in agreement with the observations obtained by a recent examination of this reaction by Ma et al. \cite{Ma}, albeit with a reasonably large shift in excitation energies, discussed below. The essential difference between our data set and that of Ma is a much larger (and different) angular range covered, 125 – 170 degrees in our work, versus 170-180 degrees in that of Ma. Therefore, the two data sets are complementary, and the new measurements performed in this work allow for an independent and more reliable evaluation of the spin-parity assignments, due to a much larger angular range coverage. \par
\begin{table*}
    \centering
    \begin{tabular}{|c|c|c|c|c|c|c|c|}
        \hline
        $J^{\pi}$ & $E_{x}$($^{14}$C) [MeV]  & $\Delta E_x$ [MeV] \cite{AMD3} & $E_{x}$($^{14}$O) [MeV] & $\Gamma_{\texttt{tot}}$ ($^{14}$C) [MeV]  & $^{14}$C $P_{l}(E_{x})$ &  $^{14}$O $P_{l}(E_{x})$ & $\Gamma_{\texttt{tot}}$ ($^{14}$O) [MeV]   \\
        \hline
        $0^{+}$ & 13.9(1) & $-$0.08 & 13.8(1) & 0.150(30) & 0.923 & 1.759 & 0.285\\
        $2^{+}$ & 14.9(1) & $-$0.24 & 14.75(1) & 0.100(20) & 0.980 & 1.619 & 0.165\\
        $4^{+}$ & 17.3(1) & $-$1.04 & 16.3(1) & 0.120(30) & 1.018 & 1.023 & 0.120\\
        $4^{+}$ & 17.3(1) & $-$0.27 & 17.0(1) & 0.120(30) & 1.018 & 1.396 & 0.165\\

        \hline
    \end{tabular}
    \caption{Conversion of previously-observed $\pi^2$ linear-chain states from $^{14}$C to $^{14}$O using mirror symmetry. The $J^{\pi}=4^{+}$ is expected as a doublet in $^{14}$C.}
    \label{tab:resonances}
\end{table*}

\begin{figure}[ht!]
\centerline{\includegraphics[width=0.5\textwidth]{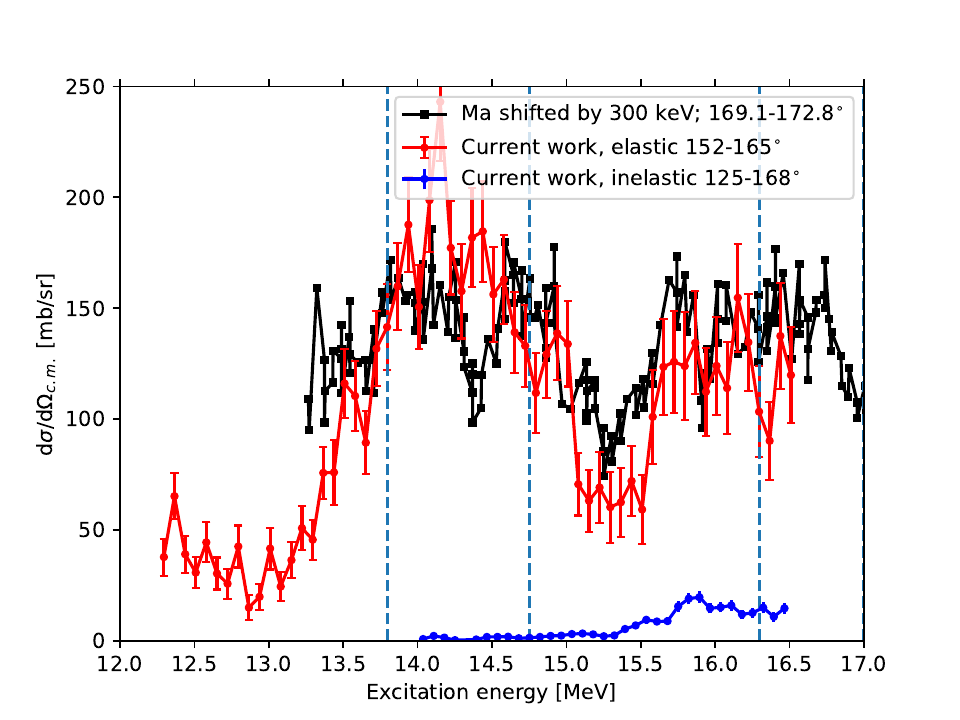}}
\caption{\label{fig:compare}Comparison of the current work for both the elastic channel for a given angular range (red) and the inelastic channel (averaged across a larger angular range due to limited statistics), against the differential cross sections obtained by Ma et al. \cite{Ma,Erratum}. Dashed vertical lines show the expect placement of the analogue $\pi^2$ rotational band states observed by Han et al.\cite{Han} as given in Table~\ref{tab:resonances}.}
\end{figure}
\section{R-Matrix comparison to the mirror system}

As stated in Sec.~\ref{sec:Introduction}, a recent paper by Han \cite{Han} claimed to see evidence of a linear-chain configuration in the mirror system, $^{14}$C. This was deduced from a series of resonances in a rotational band where the spin was obtained using the angular correlation method. A discussion of this method follows below in Sec.~\ref{sec:validity}.\par
Nonetheless, the parameters for the rotational band resonances were taken from the work of Ref.~\cite{Han}, and the reduced-widths calculated which were then used to calculate the partial widths for the mirror system $^{14}$O by assuming the same reduced-width amplitude and therefore:
\begin{eqnarray}
\gamma_{\alpha}^2 = \, ^{14}\textrm{C}(\Gamma_{\alpha}) / (2 \times \, ^{14}\textrm{C}(P_{l}(E))), \\
\, ^{14}\textrm{O}(\Gamma_{\alpha}) = 2 \, ^{14}\textrm{O}(P_{l}(E))) \gamma_{\alpha}^2, \\
\therefore \, ^{14}\textrm{O}(\Gamma_{\alpha}) = \, ^{14}\textrm{C}(\Gamma_{\alpha}) \frac{\, ^{14}\textrm{O}(P_{l}(E)))}{\, ^{14}\textrm{C}(P_{l}(E))).}
\end{eqnarray}
Here the total width measured was assumed to be equal to $\Gamma_{\alpha_0}$ as (for the majority of states) the penetrability to the first excited state is either small or energetically-forbidden. In addition, the Thomas Ehrman shifts from Ref.~\cite{AMD3} were taken, as it is claimed that these AMD results well-replicated the experimental data for linear chain states. The widths and excitation energies were taken from the experimental work rather than the theoretical work as it can be seen with our studies that the exact excitation energy is difficult to replicate and often the widths are overestimated (being a factor of 5 higher for the Hoyle state \cite{AMD4}). The highest excitation energy state at 17.0 MeV is outside of our region of study but was included to investigate the interference effect it may have with the lower-lying 16.3 MeV state. The cross section for the system of four resonances (summarized in Table~\ref{tab:resonances}) was calculated in conjunction with the resonances described by the previous work of Ma et al. \cite{Ma} as well as a series of background poles at both higher and low excitation energies which were allowed to freely vary their width. The resolution of 80 keV in the center of mass was also taken into account with the cross section being smeared by a Gaussian with this standard deviation.  For our highest statistics angle, our current data are shown in comparison to the recently-obtained data by Ma et al. \cite{Ma} (with a correction to the magnitude of the cross section published as an erratum \cite{Erratum}) in Fig.~\ref{fig:compare}. As mentioned above, it appears that there is an offset in the excitation energy between our data and those of Ma with our excitation energies approximately 300 keV higher, a sizeable difference given the inverse kinematics. One potential discrepancy between the two data sets may originate from the energy loss correction of the recoil $\alpha$-particle in the gas. Using the energy-range relations given from LISE++ and SRIM gave an almost identical result, giving confidence our energy loss was calculated correctly. When considering this discrepancy, the two data sets are otherwise in relatively good agreement with two caveats. Firstly, the setup of Ma was unable to distinguish inelastic events on an event-by-event basis and, as such, made some arguments as to why they expected the contribution to be small which seems to be in reasonable agreement with the small inelastic contribution demonstrated in Fig.~\ref{fig:compare}. Secondly, due to the fact that data are compared for a given silicon detector between two experiments, the angle for a specific excitation energy varies so as one passes through maxima or minima in the Legendre polynomials, the contribution of resonances can be increased/decreased, in specific locations large deviations between the two data sets are expected so one cannot directly compare like-for-like angles.
\begin{figure}[ht!]
\centerline{\includegraphics[width=0.5\textwidth]{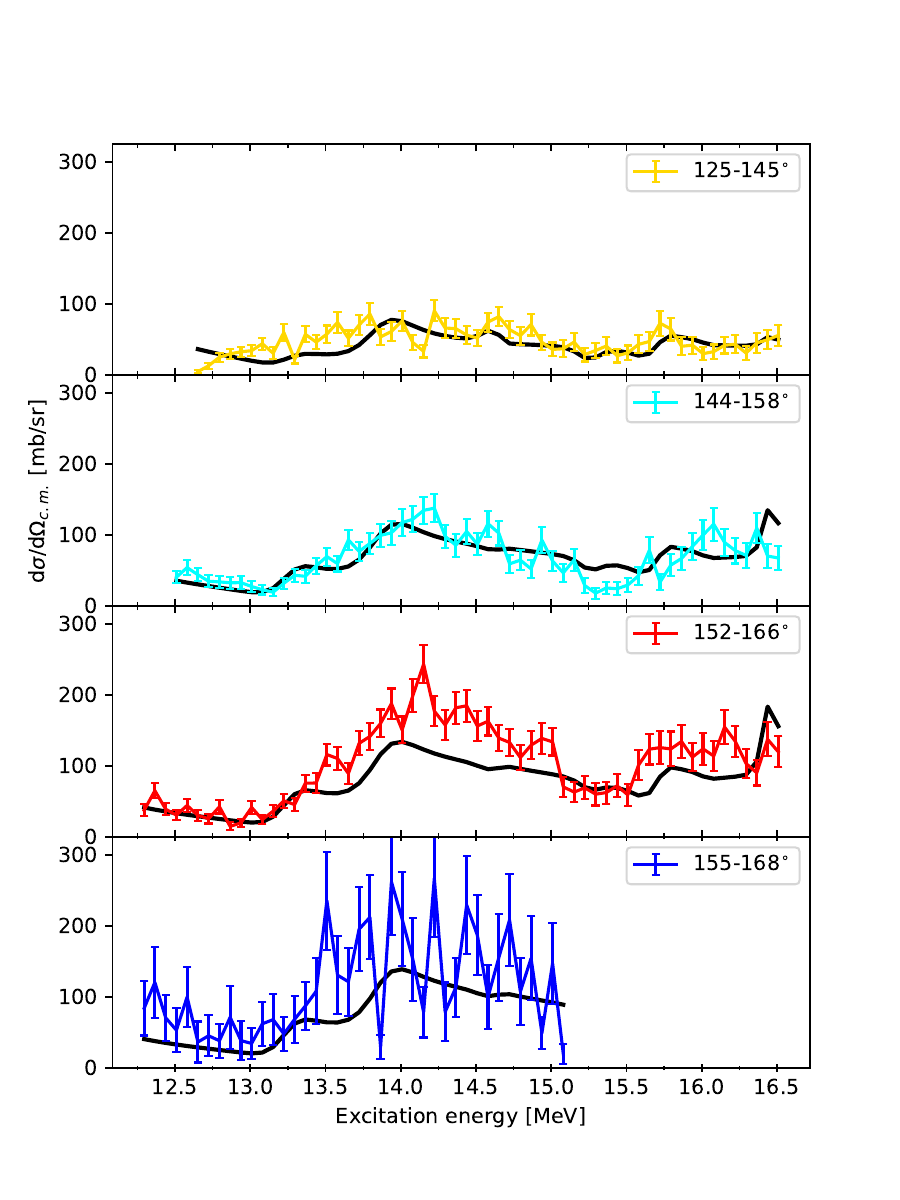}}
\caption{\label{fig:MaRMatrix}Current experimental data for four different detectors overlaid with the R-Matrix cross section as calculated with the states from Ma et al. \cite{Ma} shifted by 300 keV in excitation energy.}
\end{figure}

\begin{figure}[ht!]
\centerline{\includegraphics[width=0.5\textwidth]{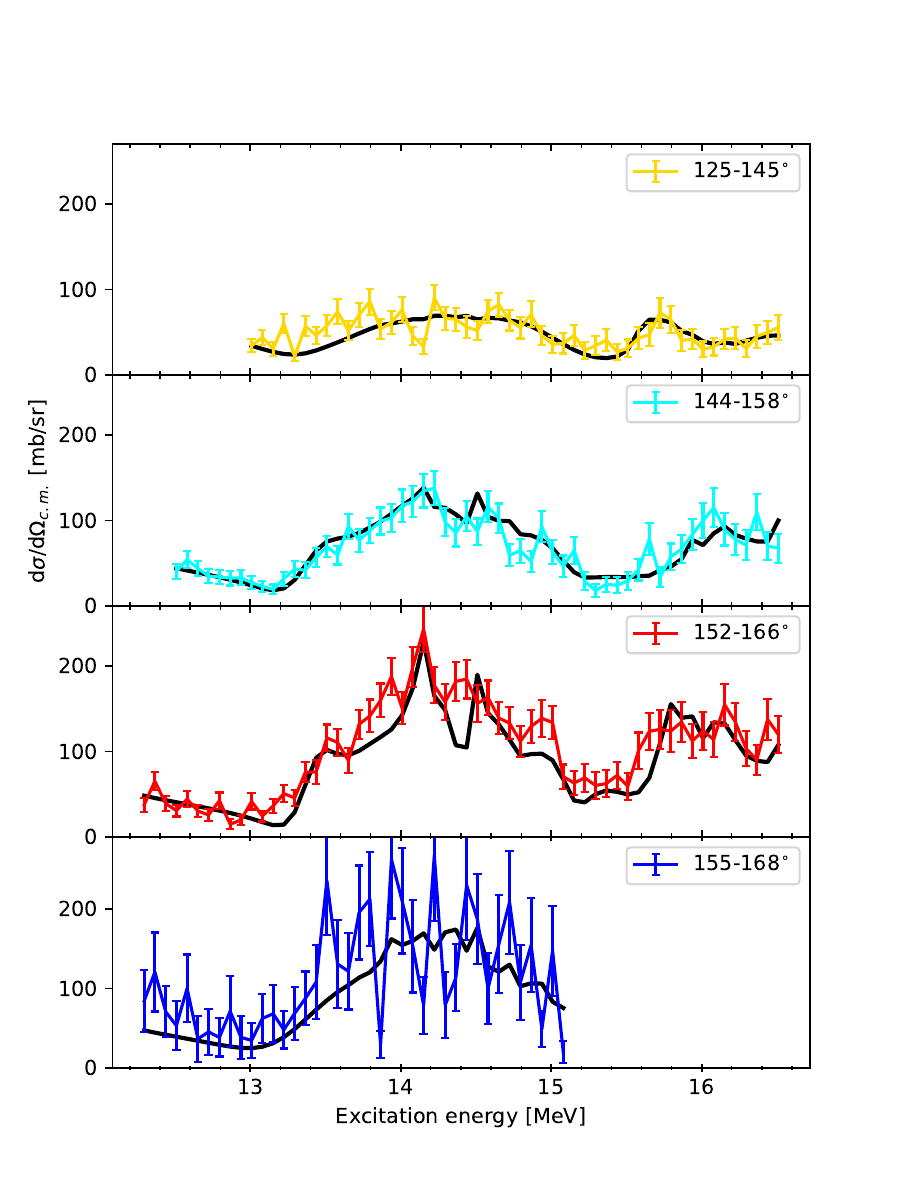}}
\caption{Current experimental data for four different detectors overlaid with the R-Matrix cross section as calculated with the states from Table \ref{tab:experimental_data}.\label{fig:mixedfit}}
\end{figure}

\begin{table*}[ht!]
    \centering
    \begin{tabular}{|c|c|c|c|c|c|c|c|c|>{\centering\arraybackslash}p{30mm}|}
        \hline
        \multicolumn{3}{|c|}{\textbf{Current Work}} & \multicolumn{3}{|c|}{\textbf{Ma} \cite{Ma}} & \multicolumn{3}{|c|}{\textbf{Literature}} & \textbf{Comment}\\
        \hline
        \textbf{$E_x$ (MeV)} & $J^{\pi}$ & $\Gamma_{\alpha_0}$\textbf{ (keV)} & \textbf{$E_x$ (MeV)}  & $J^{\pi}$ & $\Gamma_{\alpha_0}$\textbf{ (keV)} & \textbf{$E_x$ (MeV)}  & $J^{\pi}$ & $\Gamma_{\alpha_0}$\textbf{ (keV)} & \\ \hline
        13.4              & 1$^{-}$           & 443(20)              & 12.94            & 1$^{-}$           & 252(44)    & & &           & Shifted $E_x$       \\ \hline
        13.73             & 0$^{+}$           & 611(7)               & 13.43            & 0$^{+}$           & 511(143)    & & &         & Shifted $E_x$       \\ \hline
        14.2              & 3$^{-}$           & 27(3)                & 14.29            & 4$^{+}$           & 6(2)         & \begin{tabular}{c}14.15 \cite{Korkmaz}\\  14.1 \cite{Kraus} \end{tabular}   & \begin{tabular}{c}($5^{-}$)\\  4$^{+}$ \end{tabular}  &            & Different $J^{\pi}$ \\ \hline
        14.6              & 3$^{-}$           & 16(3)                &                  &              &               & 14.6 \cite{Korkmaz} & &       & Additional state \\ \hline
        15.13             & 1$^{-}$           & 100(17)              &                  &              &                   & & &   & Additional state \\ \hline
        15.46             & 2$^{+}$           & 645(70)              & 14.88            & 2$^{+}$           & 385(296)         & & &    & Shifted $E_x$ \\ \hline
        15.63             & 0$^{+}$           & 115(15)              & 15.33            & 0$^{+}$           & 295(191)       & & &      & Shifted $E_x$        \\ \hline
        15.75              & 5$^{-}$           & 45(4)                &                  &              &                   & 15.7 \cite{Kraus} & (5$^{-}$)&   & Additional state \\ \hline
        15.98             & 1$^{-}$           & 15(5)                & 15.68            & 1$^{-}$           & 8(7)            & & &     & Shifted $E_x$       \\ \hline
        16.13             & 4$^{+}$           & 14(10)               & 16.17             & (3$^{-}$)         & 13(6)          & & &           & Different $J^{\pi}$ with\newline additional inelastic\newline  width of 43(40) keV \\ \hline
        16.68             & 4$^{+}$           & 7(6)                 & 16.38            & (4$^{+}$)         & 7(2)              & & &   & Shifted $E_x$        \\ \hline
    \end{tabular}
    \caption{List of resonances in $^{14}$O used to provide an R-Matrix fit to the data based on the results of Ma \cite{Ma}. The comparison to the states presented in that work are also shown for resonances identified as the same.}
    \label{tab:experimental_data}
\end{table*}

\begin{figure}[ht!]
\centerline{\includegraphics[width=0.5\textwidth]{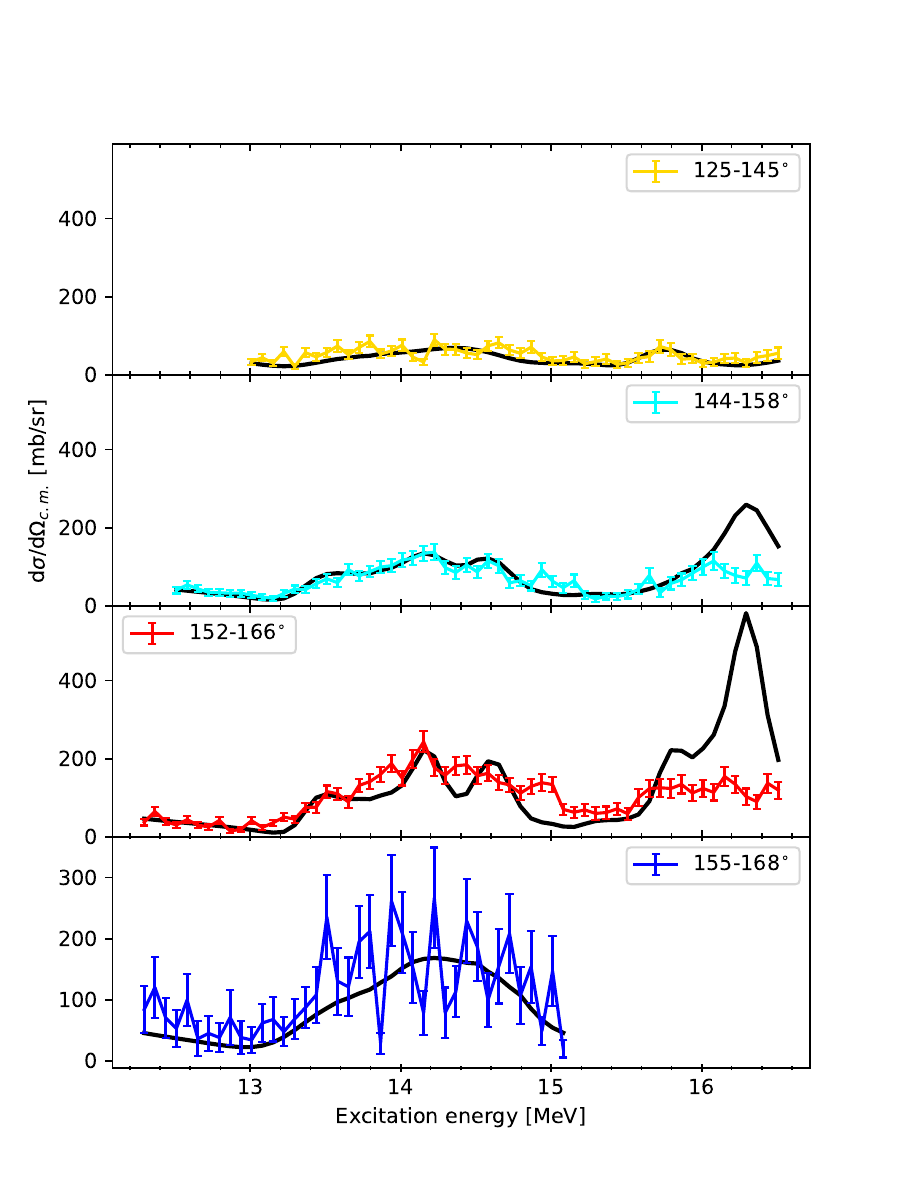}}
\caption{\label{fig:RMatrixpi2}Current experimental data for four different detectors overlaid with the R-Matrix cross section as calculated with the states from Table \ref{tab:experimental_data}, with the additional $\pi^2$ states from Table~\ref{tab:resonances}.}
\end{figure}
\begin{figure}[ht!]
\centerline{\includegraphics[width=0.5\textwidth]{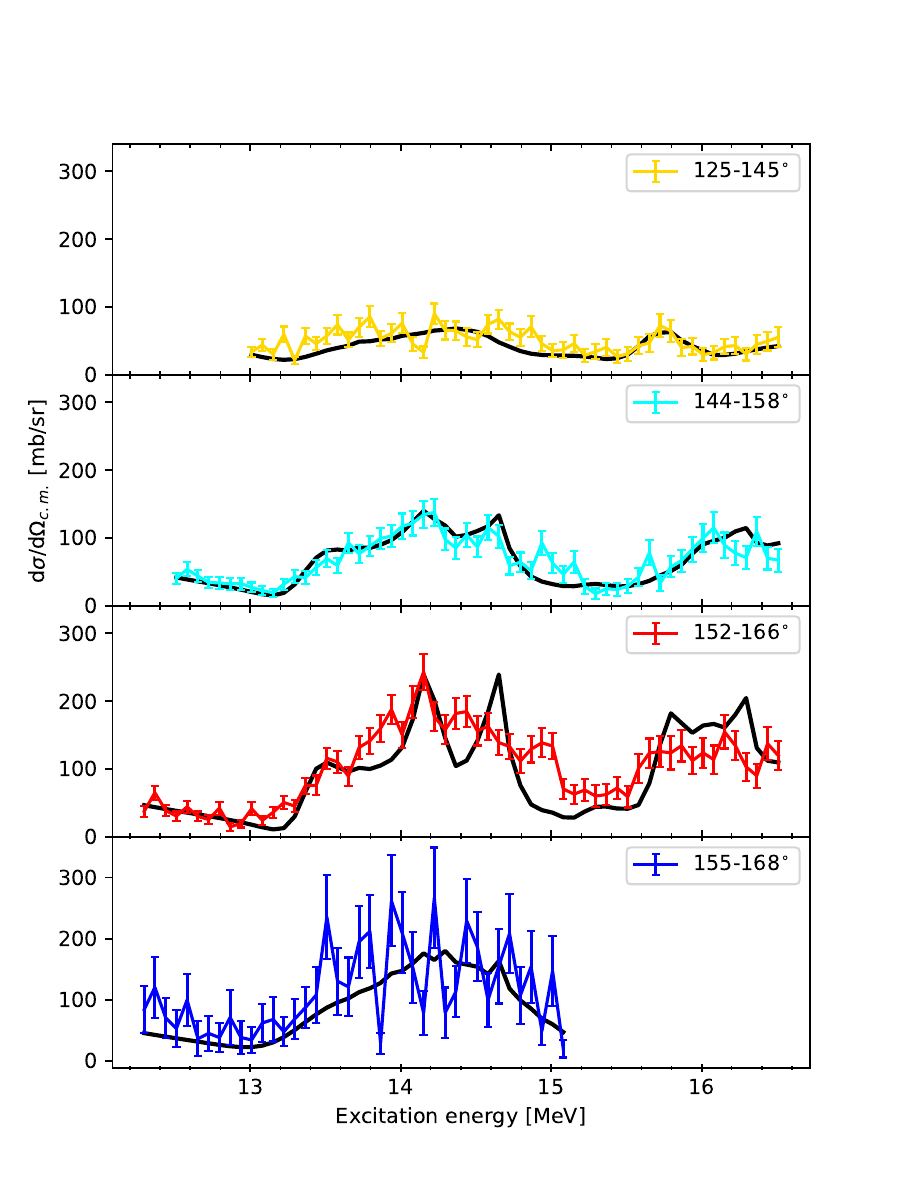}}
\caption{\label{fig:RMatrixpi2_narrow}Current experimental data for four different detectors overlaid with the R-Matrix cross section as calculated with the states from Table \ref{tab:experimental_data}, with the additional $\pi^2$ states from Table~\ref{tab:resonances}, with the 16.3 MeV resonance having $\Gamma$ = 12 keV rather than 120 keV.}
\end{figure}
\par
In order to quantitatively evaluate the role that the $\pi^2$ linear chain states would have in $^{14}$O, the resonances obtained by Ma were taken and shifted 300 keV higher in energy to match our current data and the state properties used as the basis for an R-Matrix fit. The results of our R-Matrix calculation are overlaid with our experimental data in Fig.~\ref{fig:MaRMatrix}, allowing only for variations of the background poles. For the angles closer to 180$^{\circ}$ in the c.m., the fit appears to be very good but for the outer detectors the quality of the fit signifies that the R-Matrix solution is perhaps not unique. The lack of unique allocation of resonances from different experiments is a general problem, as highlighted by the very different set of states predicted in previous studies by Fritsch \cite{Fritsch}, Freer \cite{Freer2} and Yamaguchi \cite{Yamaguchi} where there are no similar resonances, despite all reporting on $^{10}$Be$(\alpha,\alpha)$ TTIK experiments.  Attempting to fit our data \emph{de novo} also saw this lack of uniqueness where multiple fits could be established with a different collection of states. As such, we utilized the shifted parameters of Ma and attempted to make minimal modification to these resonances. Given the larger angular range of our data against those of Ma, we are more sensitive to incorrect spin-parity assignments and subsequently modified a few spin parities and also required some new resonances to describe the data further from 180$^{\circ}$ in the center of mass. A summary of these resonances is shown in Table~\ref{tab:experimental_data} with the corresponding improved R-Matrix fit being shown in Fig.~\ref{fig:mixedfit}. A much better fit to our data was demonstrated which, while we do not suggest it is an optimal classification of the set of important states in this system, both validates our data as being in agreement with those of Ma, and provides some further credence to the subset of states for which our values agree with previous data. Future effort to develop techniques using Machine Learning to quantify the uniqueness and confidence of spin assignments is therefore hugely important and is currently ongoing for future cluster studies.  \par

In Fig.~\ref{fig:RMatrixpi2}, the additional states from Table~\ref{tab:resonances} (the $^{14}$C states that have been transformed into the mirror from Han) are included into the improved AZURE2 R-Matrix calculation from Table~\ref{tab:experimental_data} to demonstrate the influence they would have on the observed cross section. The $E_x$ = 16.3 MeV $4^{+}$ state can be seen to have a very strong effect on the cross section with the expected value far exceeding that observed experimentally for the inner and middle angles. For the outer angles (top panel of Fig.~\ref{fig:RMatrixpi2}), this excitation energy corresponds roughly to where the $L=4$ Legendre polynomial goes to zero at 150$^{\circ}$ so the yield for this detector is minimal. Suffice to say, the set of states as defined in Table~\ref{tab:resonances} (specifically the 16.3 MeV state) are inconsistent with our current data. Adding additional resonances to interfere with this strong state was attempted but it is not possible to suppress the cross section sufficiently to be consistent with our data. Alternatively, by reducing the width of the 16.3 MeV state, the experimental yield will decrease and become more consistent with our data. Shown in Fig.~\ref{fig:RMatrixpi2_narrow} is the same set of states from Table~\ref{tab:resonances} but with a width one-tenth that as seen by Han et al. for the 16.3 MeV $4^{+}$ state. This can be seen to better align with our experimental data as the widths are decreased, but the expected contribution still exceeds our results. The width of the state, if it exists at this energy, is therefore safely established to be $<$10 keV which corresponds to a dimensionless width (i.e. the fraction of the Wigner limit) of 0.6$\%$ which is not a highly-clustered state and not characteristic of a $\pi^2$ linear chain state. \par
The higher-lying 4$^{+}$ state at 17 MeV is unfortunately just above the range of our data but using the data of Ma, a broad 4$^{+}$ is also not clearly visible in this region with the next strong peak corresponding to $E_x$ = 18 MeV. This was identified by Ma as a narrow ($\Gamma$ = 11 keV) $4^{+}$ state --- also corresponding to a poorly-clustered state.

\par
This leads to two likely scenarios which will be discussed initially, followed by alternative explanations. Firstly, the experimentally-observed resonances in $^{14}\mathrm{C}$ are erroneously assigned or have incorrectly determined widths. Secondly, the mirror system breaks the linear chain configuration and the linear chain rotational band states do not exist in $^{14}\mathrm{O}$ but are robust in $^{14}$C.\par

\subsection{Testing the validity of the $^{14}$C linear chain states \label{sec:validity}}
In the experimental configuration for the previous $^{14}$C work of Han et al. \cite{Han}, the incident particle of a proton has non-zero spin making an extra limitation necessary in the center-of-mass scattering angle of the excited compound nucleus, denoted in their work by $\theta^{\star}$, where  $\theta^{\star}$ should be approximately zero \cite{Curtis}. In Fig.~4 of their work, it can be seen that the majority of their data lie between 10 and 100$^{\circ}$ which does not respect this required limitation. For comparison, in the previously-cited work \cite{Curtis}, the majority of events lay within $|\theta^{\star}|<30^{\circ}$ and additionally suggests limiting to $|\theta^{\star}|<5^{\circ}$ for spin determination. \par
Furthermore, it is worth noting with comparison to the previous work of Curtis et al. \cite{Curtis}, the spin of the state measured was assigned to be $J^{\pi}=3^{-}$ but a later experiment clearly demonstrated (with all particles being zero spin) that the spin was unambiguously $J^{\pi}= 4^{+}$ \cite{Remeasure}. This specific case highlights the dangers associated with non-zero spin targets/beams and may possibly remedy the disagreement between our data in the mirror system and the result in $^{14}$C due to an incorrect spin assignment presenting an incorrect rotational band. In fact, the spin-parity of the $J^{\pi}$=4$^{+}$ is the only assignment that causes a conflict between our current data and those of Ma et al. \cite{Ma}. \par
To understand the observations better, one may turn to a previous work by Yamaguchi et al. \cite{Yamaguchi} for the $^{14}$C system where they observed a strong rotational band with similar characteristics to those of Han et al. but with different energies and widths. Crucially, they also did not observe any of the $\pi^2$ rotational band states claimed by Han while observing states with similar excitation energies and widths (e.g. assigning the $J^{\pi}=4^{+}$ state at 17.25 MeV as a potential $J^{\pi}=1^{-}$ state). The strength of the work of Yamaguchi et al. for comparison is that it was performed using TTIK and elastic scattering. There is a question in inelastic scattering experiments of how well one may populate highly-clustered states that may have a very different structure from the $^{14}$C ground state --- especially considering these are expected to be highly-deformed linear chain states. In the work of Han, they claim their advantage of high Q-value resolution allowing them to perform channel selection to remove the inelastic component of $^{10}$Be$^{star}$ makes their results more reliable, Fig.~\ref{fig:compare} however shows this contribution to be small and therefore this advantage may be of little benefit. Their subsequent poor resolution for the excitation energy spectrum ($\sim400$ keV FWHM at $E_x$ = 14.9 MeV) makes reliable width estimates of values smaller than this value difficult.
\par
To refute this possibility, one must therefore answer the question as to why a series of strong $\pi^2$-clustered states would not be observed in the current work and that of Yamaguchi et al. given their large $\Gamma_{\alpha}$.
\subsection{Possibility of mirror-symmetry breaking}
Another possibility is that the states observed by Han et al. are entirely correct, and the fact that we do not see those same states here in the mirror system is evidence of complete symmetry breaking. The expected splitting of one $J^{\pi}$ = 4$^{+}$ state in $^{14}$C into two in $^{14}$O \cite{AMD3} is already due to symmetry breaking between the two nuclei. One may look both to similar systems and theoretical calculations to test this possibility. States with an alpha-clustered basis with valence neutrons/protons are well-studied in systems like $^{9}$Be/$^{9}$B \cite{BeB9,Boronstate,Berystate,Brooks}, $^{10}$Be/$^{10}$C/$^{10}$B \cite{Kuchera,C10,Kirsebom,Charity}, and $^{18}$Ne/$^{18}$O \cite{Melina,Changbo,Stuart,Stuart2,Marina}. Remarkable agreement has been seen for these mirror systems extending up to high spin and excitation energy. For an excellent overview of this area, Ref. \cite{Yamaguchi2} more details. \par
Supporting this possibility of complete symmetry breaking is extremely difficult however as the AMD calculations that underpin the experimental work of Han seem to provide a strong prediction of the mirror system of $^{14}$C (namely $^{14}$O) having strong $\pi^2$ and $\sigma^2$ states. Disputing the robustness of this AMD prediction would also undermine the conclusions drawn by Han et al. that their experimental data match AMD calculations extremely well. In other AMD calculation work for $^{14}$C \cite{Suhara}, the orthogonality of lower-lying states is a driver for the existence/non-existence of linear chain configurations in $^{12}$C and $^{14}$C. One possible qualitative explanation is that the Thomas-Ehrmann shifts affecting the states that are not linear-chains, modifies this orthogonality condition by sufficiently changing the excitation energy between the mirror systems. Configuration mixing may also be stronger than expected meaning the resulting states do not have a large overlap with a simple $\pi^2$ molecular configuration which follows the conclusion of earlier AMD calculations by Suhara and Kanada-En'yo \cite{Suhara}.\par
This possibility however must also overcome the fact that similar states were not observed in $^{14}$C using resonant $\alpha$-scattering by Yamaguchi et al. \par

\subsection{Alternative explanations}
One may also investigate alternative reasons for the disparity observed, a few of which are briefly discussed here. Firstly, is whether the state of interest may think about whether one is being populated, but being a compound nucleus reaction into a state with a large $\alpha_0$ width, this is of course not an issue and the R-Matrix calculations demonstrate this. \par
The statistics for this current study are rather low so if there are any narrow states (where the yield in TTIK is given by the total width of the state), they may not generate a statistically-significant signal above any broad `background' states. The states of interest however are $>$100 keV in width so this effect cannot explain the inconsistencies observed, and this effect is included in the R-Matrix spectra calculated. \par
An additional test is whether the experimental resolution in the current work is sufficient to resolve the narrower states or whether the width is dominated by the reconstructed-energy resolution. While there are no strong isolated narrow resonances seen in this reaction where one can experimentally determine the resolution, simulations \cite{kinfit} and the previous work performed concurrently with this work \cite{Marina} both demonstrate the center of mass resolution is $\sigma \approx 80$ keV.

\section{Conclusion}
While our data seem consistent with the previous observations of the $J^{\pi}$ = 0$^{+}$ and $2^{+}$ states at 13.8 and 14.75 MeV excitation energy in $^{14}$O respectively, the observed spectrum at higher excitation energies for $^{14}$O seems difficult to reconcile with previous observations of $\pi^2$ linear chain states in $^{14}$C. In particular, $J^{\pi}=4^{+}$ states are seen to significantly exceed the observed cross section at several angles and inserting of additional states struggles to dampen the magnitude of the cross section. From previous experiments where the limitations of the angular correlation method with non-zero spin particles \cite{Curtis} was attempted to be overcome, the spin-parity assignment using this method was found to be erroneous \cite{Remeasure}. It is possible therefore that the $J^{\pi}=4^{+}$ member of this supposed rotational band is incorrectly assigned or that the total width is over-estimated. As such, the observed series of states may not represent a rotational band built on linear-chain states in as simplistic a way as originally postulated by earlier work. Limitations in the current experiment in terms of statistics have been discussed, as well as alternative explanations for the observed disagreement but no convincing alternate hypothesis was found. \par
Repeating the $^{14}$C inelastic excitation experiment but with a spin-zero target and also running with an $^{14}$O beam would allow for a much more robust study on the veracity of these reported states across the mirror system.
\section{Acknowledgments}
The authors would like to thank the operators at the Cyclotron Institute for delivering high-quality beams during the experiment.
This work was supported by the UK Science and Technology Facilities Council (STFC) under Grant numbers ST/P004199/1 and ST/V001043/1. This work was supported in part by the U.S. Department of Energy, Office of Science, Office of Nuclear Science under Award No. DE-FG02-93ER40773. G.V.R. also acknowledges the support of the Nuclear Solutions Institute.
\bibliography{PRCBib}

\end{document}